# Adsorption dynamics of hydrophobically modified polymers at an air-water interface

C. Trégouët[1,2], A. Mikhailovskaya[1], T. Salez[3,4], N. Pantoustier[1], P. Perrin[1], M. Reyssat*[2] and C. Monteux*[1,4]

1. Laboratoire Sciences et Ingénierie de la Matière Molle, PSL Research University, UPMC Univ Paris 06, ESPCI Paris, UMR 7615 CNRS, 10 rue Vauquelin, 75231 Paris cedex 05, France

2. UMR CNRS Gulliver 7083, ESPCI Paris, PSL Research University, 10 rue Vauquelin, 75231 Paris cedex 05, France

3. Univ. Bordeaux, CNRS, LOMA, UMR 5798, F-33405 Talence, France

4. Global Station for Soft Matter, Global Institution for Collaborative Research and Education, Hokkaido University, Sapporo, Hokkaido 060-0808, Japan

* corresponding authors : mathilde.reyssat@espci.fr; cecile.monteux@espci.fr

**Abstract**

Using surface-tension measurements, we study the brush-limited adsorption dynamics of a range of amphiphilic polymers, PAAH-$\alpha$–$C_n$ composed of a poly(acrylic acid) backbone, PAAH, grafted with a fraction $\alpha$ of alkyl moieties, containing either $n$=8 or $n$=12 carbon atoms, at pH conditions where the PAAH backbone is not charged. At short times, the surface tension decreases more sharply as the degree of grafting increases while at long times, the adsorption dynamics becomes logarithmic in time and is slower as the degree of grafting increases. This logarithmic behavior at long times indicates the building of a free-energy barrier which grows over time. To account for the observed surface tension evolution with the degree of grafting we propose a scenario, where the free-energy barrier results from both the deformation of the incoming polymer coils and the deformation of the adsorbed brush. Our model involves only two fitting parameters, the monomer size and the area needed for one molecule during adsorption and is in agreement with the experimental data. We obtain a reasonable value for the monomer size and find an area per adsorbed polymer chain of the order of 1nm$^2$, showing that the polymer chains are strongly stretched as they adsorb.

## Introduction

Polymers are widely used as interface stabilizers, emulsifiers, foamers or suspension dispersants. Adsorbed at liquid/liquid or solid/liquid interfaces, polymers form loops, which provide a steric protection for droplets or particles against coalescence or flocculation respectively. Amphiphilic polymers, such as random and block copolymers or polymers grafted with hydrophobic anchors adsorb more strongly to interfaces than homopolymers. The structure of adsorbed amphiphilic polymer layers has been the object of theoretical and experimental research in the 90s[1]. For block copolymers, it was predicted that the hydrophobic parts form 2D coils while the hydrophilic parts form 3D coils swollen in the bulk solution[2,3]. At high surface polymer concentration, the hydrophilic parts stretch perpendicularly to the interface thus forming a brush[2–9]. Brushes are also expected for other amphiphilic polymers such as telechelic[10], end-functionnalized[11] or hydrophobically modified polymers[12–16], for which hydrophobic grafts adsorb to the interface and hydrophilic parts stretch into the bulk solution. The adsorption dynamics of amphiphilic polymers[17–24] has attracted much less attention than their structure, even though it is relevant in foaming or emulsification processes as well as for foam and emulsion stability[25,26]. A fine understanding of the influence of the polymer architecture on the adsorption dynamics is however required for desired applications.

In this article, we investigate the adsorption dynamics of a series of hydrophobically modified polymer molecules, denoted PAAH-$\alpha$–$C_n$, composed of a PAAH backbone, poly (acrylic acid), covalently and randomly grafted with hydrophobic alkyl anchors, $C_n$, of varying length $n$ and grafting degree $\alpha$ at pH=3, where the chains are uncharged. These polymers, with a comb-like architecture, are known to be very efficient foam and emulsion stabilizers[23,27–29]. Their chemical architecture - molar mass, number of grafts, degree of ionization and length of alkyl chain - can be easily modified. Therefore they are good candidates to investigate how chemical parameters influence the adsorption dynamics of amphiphilic polymers. We focus here on long time evolution of the interfacial tension, which decreases in a logarithmic manner with time, a process which is slower for higher degrees of grafting. Several models have been proposed to account for a logarithmic evolution of the interfacial tension but none of them considers the case of statistically grafted polymers. The Ward and Tordai model[30], designed for surfactants or globular proteins, takes into account the surface pressure to overcome to adsorb a new molecule, which grows over time. The Ligoure and Leibler[11], and the Johner and Joanny[5] models describe the adsorption of amphiphilic end-grafted polymers and diblock copolymers, respectively. In their cases, the free-energy barrier is of entropic origin, due to the deformation of one adsorbing chain which stretches to be able to enter the adsorbed brush and the deformation energy increases as the brush becomes denser over time. While interesting and valid in its own framework of application, each of these models cannot be directly used to describe our statistically grafted polymers. In fact, our polymers adsorb their grafts, with hydrophilic loops in between the grafts whose length depends on the grafting density[13]. Nevertheless, by combining the ingredients above, i.e the deformation of the brush and of the adsorbing molecules, we develop a novel model that captures well the experimental data, and thus offers a preliminary coarse-grained picture of the possible mechanism at play in amphiphilic polymer adsorption.

**Materials**

We use a series of hydrophobically modified polymers, denoted PAAH-$\alpha$-$C_n$, composed of a PAAH, poly(acrylic acid) backbone, of molar mass $M_W$=100 000 g/mol, of Poly Dispersity Index PDI=2, covalently grafted with a mole fraction $\alpha$ of hydrophobic alkyl chains comprising $n$ carbon atoms. The PAAH backbone is provided by Polysciences Inc. and the grafting reaction was done according to Wang *et al.* [31]. In our work, $\alpha$ ranges between 0.5/100 and 5/100 and $n$ = 8 or 12. We keep the degree of grafting low enough to ensure a full solubility of the chains in water in acidic conditions. An illustration of the comb-like chain architecture is given in Fig. 1. We find (shown in SI) that the viscosity of the PAAH-$\alpha$–$C_n$ solutions is slightly lower than that of pure PAAH backbones, indicating some possible intramolecular association of the hydrophobic grafts. Using Dynamic Light Scattering, DLS, we find that the hydrodynamic radius of the grafted PAAH-$\alpha$-$C_n$ series is of the order of 10 nm.

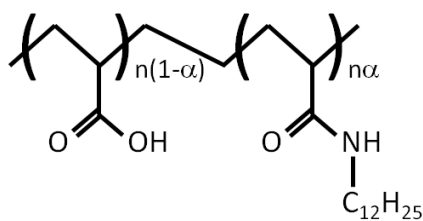

Figure 1: *Chemical structure of the PAAH-$\alpha$–$C_n$ chain, in the n=12 case.*

To prepare the polymer solutions, the chains are dissolved in deionized water during 60 hours. The polymer weight fraction is fixed at 1%. The pH is adjusted to pH=3 with a few milliliters of molar hydrochloric acid solution, to obtain conditions in which the PAAH chains do not bear any negative charge. The solutions are then filtered through 0.45 µm polyvinylidene difluoride (PVDF) membranes, before being stored at 4 °C. For each experiment the solutions are kept at ambient temperature for 12 hours prior to the experiment. The solutions are used within seven days after the polymer dissolution.

**Methods**

To probe the adsorption dynamics of the polymer chains, we use the pendant-drop method (Tracker apparatus, Teclis, France), which enables the measurement of the surface tension as a function of time. Details concerning the principle of this technique can be found in an article from Rotenberg et al.[32]. Briefly, a fresh millimetric air bubble is formed in the polymer solution and the surface tension is then obtained as a function of time, thanks to the dynamical analysis of the bubble shape and the Young-Laplace law. In the following we call $t_i=0$ the time at which the first surface tension measurement is made, which corresponds to the time at which the bubble formation process, which takes roughly three seconds, is completed.

The temperature is maintained constant at 20 °C.

**Results and discussion**

The interfacial tension $\gamma$ of the air bubble in the polymer solution is recorded as a function of time, and presented in Fig. 2a using linear scales and 2b using a semi-log plot. As explained above in the experimental section, the time $t_i=0$ is the time at which the bubble formation process is completed. During the bubble formation, which takes roughly 3 seconds, the polymer molecules adsorb to the interface by diffusion, which explains why the surface tension is not equal to that of pure water with air (72 mN/m) when the measurement is started. Indeed the typical time needed to reach a surface excess $\Gamma$ for a bulk concentration $C_0$ writes $\tau \sim (\Gamma/C_0)^2/D \sim 1$ ms for $\Gamma=1$mg/m$^2$, $C_0 = 10^4$g/m$^3$ and $D = 10^{-11}$ m$^2$/s.

At longer times, the interfacial tension decreases, indicating that the surface excess at the interface increases. For the grafted polymers, the interfacial tension decreases faster and to a larger extent than for non-grafted

PAAH. Indeed the interfacial tension between air and the PAAH solution is very close to the surface tension of water, $\gamma_0$ = 72 mN/m, which means that the adsorption of PAAH chains at the interface is weak. We would like to point out that despite the impression given by the logarithmic scale in Figure 2b, there is no real plateau in the interfacial tension evolution as seen in Figure 2a in linear scales: the interfacial tension decreases steeply at short time. Moreover Figure 2 shows that increasing the grafting density enables to decrease more sharply the interfacial tension at short times. Indeed, the driving mechanism for the adsorption of the alkyl grafts at the air-water interface is the enthalpic gain as the hydrophobic grafts come in contact with air. Hence, when the total concentration of grafts increases, the surface chemical potential equilibrates with its bulk counterpart and more grafts adsorb to the interface. We note that increasing the graft length from C8 to C12 leads to a weak decrease of the surface tension at short time, suggesting that the actual number of adsorbed grafts may be close for both types of polymers. At longer times the adsorption kinetics slows down and becomes logarithmic-like in time (see Fig. 2b), consistently with an adsorption barrier which grows with the surface concentration[5,11,30]. Surprisingly, the adsorption dynamics becomes slower for the polymers with the highest grafting degree $\alpha$, possibly indicating an increase in rearrangement constraints when multiple adsorption sites are present per chain in a densely adsorbed layer. These results differ from those of Millet et al.[12], who investigated a similar series of PAANa-$\alpha$-$C_n$ in their charged form, and found that the adsorption dynamics at long time does not depend on the grafting density.

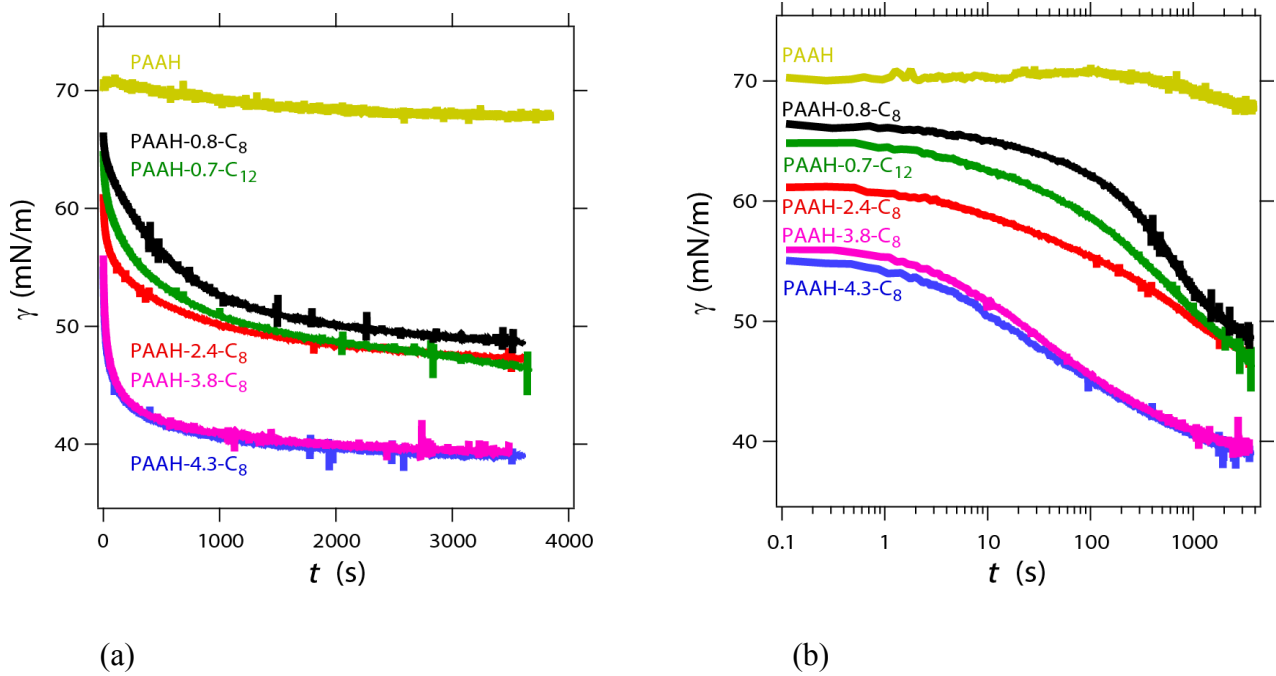

*Figure 2: Air - polymer solution interfacial tension versus time during adsorption of PAAH-α-$C_n$ chains on an air bubble. (a): linear scale. (b): semi-logarithmic scale. The plateau visible on the logarithmic scale is due to the logarithmic representation.*

Several models in the literature have been proposed to account for logarithmic adsorption kinetics. The kinetic Ward and Tordai (WT)[30] model was developed for surfactants or globular proteins. It considers that for a molecule to adsorb a mechanical work has to be paid to overcome the surface pressure of the layer, which reads

$$E_{WT} = \mathcal{A}\Pi = \mathcal{A}(\gamma_0 - \gamma), \quad (1)$$

where $\mathcal{A}$ is the area of the interface needed for a molecule to adsorb, $\Pi$ the interfacial pressure, $\gamma_0$ the surface tension of pure water with air and $\gamma$ the interfacial tension of the polymer solution with air. At long times, the free-energy barrier increases as the surface pressure increases.

The Johner and Joanny (JJ)[5], and Ligoure and Leibler (LL)[11] models have considered the adsorption of amphiphilic diblock copolymers or end-functionalized polymers respectively. The increasing entropic barrier considered in these models is due to the increasing stretching of the polymer chains as they penetrate the denser adsorbed brush.

Unfortunately, none of these models corresponds to the adsorption of statistically grafted amphiphilic polymers, which adsorbed structure is very different from the one considered in these models. Diblock copolymers or end-grafted polymers contain only one adsorbing anchor per chain and hence do not form loops, while in our case, the molecules contain several hydrophobic anchors that stick to the interface while the acrylic acid monomer units between the grafts form stretched loops in the solution. Such a structure was proposed by Millet *et al*[13] based on X-ray reflectivity measurements in the case of a PAAH-1-C$_{12}$ ($M_W$=120000 g/mol).

In the following, we propose a simple scenario which can account for the observed variation of the interfacial tension with the degree of grafting. As a first approach, we assume that all the hydrophobic grafts of a given chain present at the interface adsorb to the latter, consistently with Millet *et al.*'s X-ray results[13]. Moreover we assume that the PAAH units do not adsorb to the interface, consistently with the very low surface activity of the non-grafted PAAH. In our idealized picture we consider that the adsorption process is kinetically limited by two free-energy barriers: i) a WT-like enthalpic barrier, $E_{WT}$ caused by the surface pressure to overcome in order to adsorb one molecule; and ii) a LL- or JJ-like entropic barrier $E_{def}$ due to the stretching of the chains as they penetrate the brush. In addition, we assume that the dynamics of the potential polymer micelles in the bulk does not influence our regime of brush-limited adsorption, as predicted by Johner and Joanny[5]. In fact, Petit-Agnely *et al.*[33] studied experimentally the bulk association of PAAH-8-C$_8$ and found that for a concentration similar to the one we use here, the percentage of chains involved in the micelles is negligible, around 10%, and that the dissociation time of the micelles is of the order of 10 ms. Hence, the dissociation of the micelles is too fast to influence the adsorption dynamics of the PAAH-α–C$_8$.

In addition, as stated above, the time needed for the chains to diffuse to an empty interface and saturate it is of the order of 1ms, which allows us to consider that the brush is already formed as the first interfacial tension measurement is made. Furthermore it was shown by Millet *et al.*[12] that this brush regime is observed as soon as the interfacial tension drops by approximately 3 mN/m with respect to the pure air-water interface, a criterion which is verified for all our data points corresponding to the grafted polymers (see Fig. 2).

Far from equilibrium, we neglect the desorption flux and consider only the adsorption flux, limited by the two free-energy barriers, which reads

$$\frac{d\Gamma}{dt} = k_0\, C_0\, e^{-\frac{\Pi \mathcal{A} + E_{\text{def}}}{k_B T}}, \qquad (2)$$

with $\Gamma$ the number of adsorbed polymer chains per unit area, $E_{\text{def}}$ the entropic barrier associated with the deformation of the chains to enter the brush, $C_0$ the bulk polymer concentration, $k_0$ an adsorption constant (with unit of a speed), $k_B$ the Boltzmann constant and $T$ the temperature. Assuming a simple molecular diffusion, the adsorption constant $k_0$ is expected to be typically given by[5]

$$k_0 = \frac{R_H}{\tau_0}, \qquad (3)$$

where $R_H$ is the hydrodynamic radius, $\tau_0 \approx \frac{R_H^2}{D}$ is a microscopic diffusion time, and $D = \frac{k_B T}{6\pi \eta R_H}$ the diffusion coefficient of the polymer coils diluted in a good solvent of viscosity $\eta$. Alternatively, one could consider the viscosity of the brush to be more relevant than the viscosity of the solvent, but the viscosity has a small influence on the results.

Furthermore we assume that the molecules approach the interface as coils and that the barrier associated with the stretching of the coils as they enter the adsorbed layer can be estimated as the entropic cost required for a solvated chain of Flory radius $R_F = b\, N^{3/5}$ (where $b$ is the monomer length and $N$ the number of monomers per chain) to enter a cylindrical pore of cross section $\mathcal{A}$[34]:

$$E_{\text{def}} = \lambda\, k_B T \left(\frac{R_F}{d}\right)^{\frac{5}{3}} = \lambda\, k_B T \left(\frac{\sqrt{\pi}}{2}\right)^{\frac{5}{3}} N \left(\frac{b^2}{\mathcal{A}}\right)^{\frac{5}{6}}, \qquad (4)$$

where $d$ is the pore diameter and $\lambda$ a dimensionless prefactor of order unity that will be discussed later. We consider for now that $\lambda=1$.

In the following we will take $R_H \sim R_F \sim b N^{3/5}$. Indeed using Dynamic Light Scattering we found $R_H = 10$ nm, which is close to the value of $R_F \sim b N^{3/5} \sim$ 8 to 12 nm with $N \sim$ 500 to 1000 and $b = 2$ Å.

In a first approach, we will assume that the interfacial area $\mathcal{A}$ needed for one chain to adsorb is constant over time. Indeed the area $\mathcal{A}$ is the minimal surface area that an adsorbing chain needs to clear in order to adsorb but does not necessarily correspond to the final area that the chain will occupy at the interface as the molecules may rearrange over time to put more monomers or grafts in contact with the interface.

Then, to integrate Eq. (2) and obtain γ(t), we need a relation between the surface pressure $\Pi$ and the surface concentration $\Gamma$. According to Aguié-Béghin et al.[2] the surface pressure of a copolymer layer in a regime where the hydrophobic parts are adsorbed to the air/water interface and the hydrophilic parts form stretched loops in solution reads

$$\Pi = \frac{k_B T N \Gamma}{N_B}, \quad (5)$$

where $N_B$ is the number of monomers per loop. In our case we assume that a polymer chain composed of $N$ monomers typically contains $\alpha N$ grafts, and $\alpha N$ loops, which contain $1/\alpha$ monomers each. Therefore Equation (5) becomes

$$\Pi = k_B T \alpha N \Gamma, \quad (6)$$

Injecting Eq. (6) into Eq (2) leads to the following differential equation

$$\frac{d\Pi}{dt} \cdot e^{\frac{\Pi \mathcal{A}}{k_B T}} = k_B T\ \alpha N\ C_0 k_0\ e^{-\frac{E_{def}}{k_B T}}, \quad (7).$$

Integrating Eq. (7) between $t_i$ and $t$ leads to

$$\gamma(t) = \gamma(t_i) - \frac{k_B T}{\mathcal{A}} \ln\left(1 + \frac{t-t_i}{\tau}\right), \quad (8)$$

with $t_i$ the time at which the first measurement is performed and

$$\tau = \frac{1}{\mathcal{A} N \alpha C_0 k_0}\ e^{\frac{\mathcal{A}\Pi(t_i) + E_{def}}{k_B T}} \quad (9)$$

We fit the data of Fig. 2 with Eq. (8), using $\mathcal{A}$ and $b$ as adjustable parameters. As shown in Fig. 3a, the model captures well the experimental data. We find that $b \sim 2$ Å independently of the grafting degree. This value of $b$ is close to the value given in the literature (2.5 Å)[35-37]. The area $\mathcal{A}$ increases from 1 nm² for the PAAH-0.7-$C_{12}$ and PAAH-0.8-$C_8$ to roughly 2 nm² for the PAAH-2.4-$C_8$ and PAAH-4.3-$C_8$. These measured areas are much smaller than the cross section of our polymer coil (S ~ $\pi R_F^2$ ~ 300 nm²), which is consistent with the fact that the chains are strongly stretched. The area $\mathcal{A}$ tends to be lower for lower grafting degree. Assuming that the thickness of the adsorbed layers scales like the radius $R_B$ of the loops situated between the hydrophobic grafts, $h \sim R_B \sim bN_B^{3/5}$ (with $N_B$ the number of monomers per loop), we deduce that the thickness of the brushes decreases with the degree of grafting, from 6 nm for the PAAH-0.7-$C_8$ to 1 nm for PAAH-4.3-$C_8$. Hence the adsorbing chains tend to stretch more in the case of thicker brushes, which provide a stronger steric repulsion against incoming coils.

As the chains strongly stretch upon adsorption, we suggest that they may not adsorb all their grafts at a time, but in a sequential process. The next question we thus address is: can we estimate the number of grafts that adsorb to the interface when a section $\mathcal{A}$ of the brush is cleared?

From surface tension measurements of surfactants with a $C_8$ hydrophobic tail[38], we know that the area per $C_8$ tail for a saturated interface is of the order of 0.6 nm². Hence the area of 1 nm² obtained from our fits for PAAH-0.8-$C_8$ is consistent with only one graft adsorbing at a time. For the most grafted chains, the area $\mathcal{A}$ deduced from the fits is close to 2 nm², which would correspond to a value of three grafts adsorbing simultaneously. Cooperative adsorption of the grafts, obtained for more highly grafted polymers may be favored by stronger intramolecular attraction between the grafts in the adsorbing coils.

As a final remark, we recall that we have chosen $\lambda = 1$ in Eq. (4). Considering a slightly different value for $\lambda$ would be equivalent to changing slightly the fitted value of $b$ (as $b \sim \lambda^{-3/5}$ due to the dominant exponential part). Nevertheless, $b$ would remain independent of $\alpha$ and its order of magnitude would not vary.

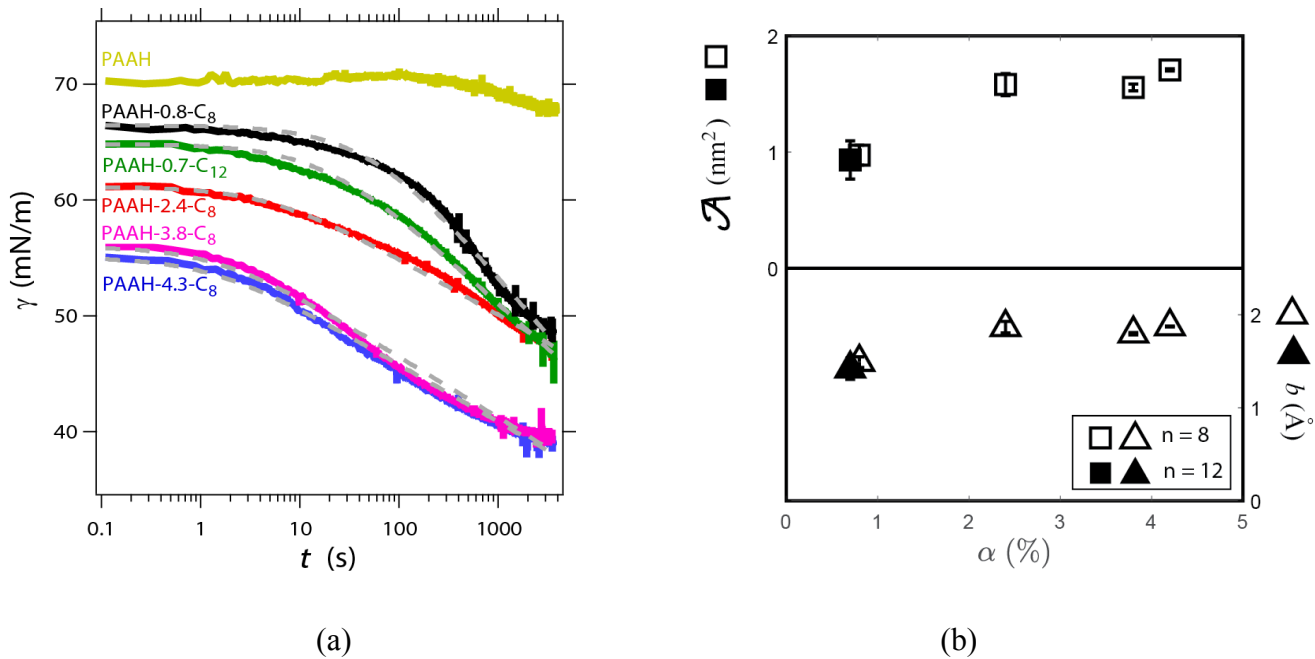

*Figure 3. (a) Air - polymer solution interfacial tension versus time obtained experimentally (solid lines, see Fig. 2) and fits using Eq. (8) (dashed lines). (b) Values of the adjustable parameters $\mathcal{A}$ and $b$ that optimize the fitting of the experimental data. n is the number of carbon atoms per alkyl grafts and α is the grafting degree of the polymer chains. Error bars represents the standard deviation calculated on 23, 20, 21, 17, 8 experiments respectively for α=0.7, 0.8, 2.4, 3.8, 4.2.*

**Conclusion**

In this article, using classical surface-tension measurements, we studied the adsorption dynamics of a range of amphiphilic polymers during the brush-limited regime of adsorption. We measured a lower surface tension with increasing grafting degree at early times, as well as a logarithmic temporal decrease at long times, which becomes slower as the degree of grafting increases. To rationalize our findings, we develop a minimal analysis which takes into account both the stretching of the adsorbing polymer chains and the deformation of the adsorbed layer. Despite some important simplifications, to be refined in future, this model provides a preliminary coarse-grained picture that agrees well with the experimental data. From the two adjustable parameters, we recover the expected value for the monomer size and we find that the adsorbed area is on the order of 2 nm$^2$, thus suggesting that the adsorbed polymer chains are highly compacted. These tunable amphiphilic surfactant polymers raise a number of fascinating questions, and might allow for a precise design of the polymer architecture in order to stabilize foams and emulsions, among other possible applications. Further experiments involving reflectivity methods would certainly help obtaining a better characterization of the structure of the layers over time[9]. In addition the model can be further refined by taking into account the possible effect of polydispersity, sequential adsorption of the grafts or desorption of the grafts at long times, as well as the fine structure of the brush and its dependency on the osmotic pressure.

**Acknowledgements**

This work was financially supported by ANR JCJC INTERPOL and ITN SOMATAI. The authors thank Jean-François Joanny, Ludwik Leibler and Julien Dupré de Baubigny for fruitful discussions, Mohamed Hanafi for GPC and DLS measurements, as well as the Global Station for Soft Matter, a project of Global Institution for Collaborative Research and Education at Hokkaido University.

**GRAPHICAL ABSTRACT**

Dynamic surface-tension measurements are used to study the adsorption dynamics of a series of hydrophobically modified polymers at the air-water interface. The adsorption dynamics is slow and follows a logarithmic behavior at large times, which indicates an adsorption barrier that grows over time. This dynamics is well described by a model which accounts for both the deformation of the incoming polymer coils and the deformation of the adsorbed brush at the interface.

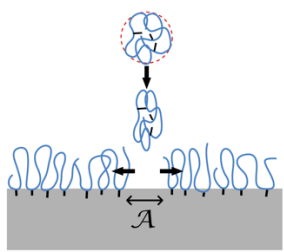